\documentclass[11pt]{article}
\usepackage{moriond,epsfig}
\usepackage{psfrag}

\def\be{\begin{equation}}
\def\ee{\end{equation}}
\def\bea{\begin{eqnarray}}
\def\eea{\end{eqnarray}}

\newcommand{\GeV}{{\ensuremath\mathrm{\;GeV}}}
\newcommand{\ifb}{{\ensuremath\rm\;fb^{-1}}}
\newcommand{\text}[1]{{\ensuremath\rm #1}}

\begin{document}
\vspace*{4cm}
\title{Extended Scalar Sector and Fat Jets}

\author{ M.~Rauch \footnote{for the SFitter collaboration} }

\address{Institute for Theoretical Physics, Karlsruhe Institute of Technology, Karlsruhe, Germany}

\maketitle\abstracts{
After a discovery of the Higgs boson the next question is what are its
couplings. At the LHC there should be many observable channels which can
be exploited to measure the relevant parameters in the Higgs sector.
Using the SFitter framework we map these measurements onto the parameter
space of a weak-scale effective theory with free Higgs boson couplings.
Our analysis benefits from the parameter determination tools and the
error treatment used in new--physics searches, to study individual
parameters and their error bars as well as parameter correlations.
A special focus we will put on recent analyses using jet
substructure techniques.
}

\section{Introduction}
Understanding electro-weak symmetry breaking is one of the main goals of
the LHC. In the Standard Model (SM) this is done spontaneously, and
achieved by an $SU(2)$ doublet, the Higgs field~\cite{higgs,reviews},
which obtains a
vacuum expectation value (vev). Three degrees of freedom become the
longitudinal modes of $W$ and $Z$ bosons, while one remains as a
physical scalar, the Higgs boson. The gauge boson masses arise from the
kinetic term of the Higgs field in the Lagrangian $(D_\mu \Phi)^\dagger
(D^\mu \Phi)$, which leads to $WWHH$ and $ZZHH$ terms, and replacing the
Higgs field by its vev then to the masses. As these and the
electromagnetic coupling are measured, this allows to determine the vev
$v=246 \GeV$ before actually observing the Higgs boson.
Fermion masses we obtain from terms $y_f H\bar{\Psi}\Psi$. The Yukawa
couplings $y_f$ are a priori free parameters in the Lagrangian, but can
be traded for the known fermion masses $m_f = y_f \cdot v$. This also
means that the coupling of the Higgs to fermions scales as the
fermion masses.

The mass of the Higgs boson is the only unknown parameter in
the SM. Electroweak precision tests tell us that it should be relatively
light, just above the limit of $114.4 \GeV$ from direct searches at
LEP~\cite{Alcaraz:2007ri,Collaboration:2008ub}. Therefore we can use the
theoretically predicted coupling values and test
this~\cite{duehrssen,Lafaye:2009vr} against future LHC
measurements~\cite{atlas_tdr,cms_tdr}.
Thereby we assume that the discrete quantum numbers, like spin and $CP$
structure, of the boson are known~\cite{wbf_vertex} and identical to the
SM.  Still there are many possible models which can generate such
deviations.
Examples include simple extensions like adding another Higgs doublet,
which is e.g.\ required in supersymmetry~\cite{susy,m_h}, or composite
models~\cite{Espinosa:2010vn}, where the Higgs emerges as a
pseudo-Goldstone boson from a strongly-interacting sector.

A correct treatment of errors is a crucial part for LHC parameter
studies. Statistical errors from event counting are of the Poisson type.
Systematic errors are correlated between individual measurements, so we
need to include the full correlation matrix. Theory errors are best
described as box-shaped using the RFit scheme~\cite{ckmfitter}. Using
the SFitter tool~\cite{sfitter} we can construct a fully-dimensional
log-likelihood map of the parameter space. If we want to ask more
specific questions, this must be reduced to lower-dimensional ones.
Bayesian marginalization and Frequentist profile likelihoods are the two
techniques we use, where the
choice depends on the specific question we want to ask.

In this article we will first review the LHC measurement channels which
enter into our analysis. A particular emphasis we will put on the
recently developed method of using jet substructure
techniques~\cite{subjet}.
Then we will present and discuss our results for Standard Model data. We
focus on a Higgs boson of $m_H = 120 \GeV$, which is the preferred
region from electro-weak precision data. 

\section{Measurements}
\begin{table}
\caption{Signatures included in our analysis for a Higgs mass of
  120~GeV. The Standard Model event numbers for $30~\ifb$ include
  cuts~\protect\cite{duehrssennote,Lafaye:2009vr}. The factor after the
  background rates describes how many events are used to extrapolate into
  the signal region. The last two columns give the one-sigma experimental
  and theory error bars on the signal. Table taken from
  Ref.~\protect\cite{Lafaye:2009vr}}
~\\
\begin{tabular}{|l|l||r|r@{ }l|r||r|r|}
\hline
 production & decay & 
 $S+B$ & 
 \multicolumn{2}{|r|}{$B$ }& 
 $S$ & 
 $\Delta S^\text{(exp)}$ &  
 $\Delta S^\text{(theo)}$ \\ \hline
 $gg \to H$ & $ZZ$ & 
  13.4 & 6.6 & ($\times$ 5) & 6.8 & 3.9 & 0.8 \\
 $qqH$ & $ZZ$ & 
  1.0  & 0.2 & ($\times$ 5) & 0.8 & 1.0 & 0.1 \\
 $gg \to H$ & $WW$ & 
  1019.5 & 882.8 & ($\times$ 1) & 136.7 & 63.4 & 18.2 \\
 $qqH$ & $WW$ & 
  59.4 & 37.5 & ($\times$ 1) & 21.9 & 10.2 & 1.7 \\
 $t\bar{t}H$ & $WW (3 \ell)$ & 
  23.9 & 21.2 & ($\times$ 1) & 2.7 & 6.8 & 0.4 \\
 $t\bar{t}H$ & $WW (2 \ell)$ & 
  24.0 & 19.6 & ($\times$ 1) & 4.4 & 6.7 & 0.6 \\
 inclusive & $\gamma\gamma$ & 
  12205.0 & 11820.0 & ($\times$ 10) & 385.0 & 164.9 & 44.5 \\
 $qqH$ & $\gamma\gamma$ & 
  38.7 & 26.7 & ($\times$ 10) & 12.0 & 6.5 & 0.9 \\
 $t\bar{t}H$ & $\gamma\gamma$ & 
  2.1 & 0.4 & ($\times$ 10) & 1.7 & 1.5 & 0.2 \\
 $WH$ & $\gamma\gamma$ & 
  2.4 & 0.4 & ($\times$ 10) & 2.0 & 1.6 & 0.1 \\
 $ZH$ & $\gamma\gamma$ & 
  1.1 & 0.7 & ($\times$ 10) & 0.4 & 1.1 & 0.1 \\
 $qqH$ & $\tau\tau (2 \ell)$ & 
  26.3 & 10.2 & ($\times$ 2) & 16.1 & 5.8 & 1.2 \\
 $qqH$ & $\tau\tau (1 \ell)$ & 
  29.6 & 11.6 & ($\times$ 2) & 18.0 & 6.6 & 1.3 \\
 $t\bar{t}H$ & $b\bar{b}$ & 
  244.5 & 219.0 & ($\times$ 1) & 25.5 & 31.2 & 3.6 \\
 $WH/ZH$ & $b\bar{b}$ & 
 228.6 & 180.0 & ($\times$ 1) & 48.6 & 20.7 & 4.0 \\\hline
\end{tabular}
\label{tab:channels}
\end{table}

At the LHC there are four main production modes of the Higgs boson:
gluon fusion, weak-boson-fusion, associated production with vector
bosons and associated production with a top-quark pair~\cite{Djouadi:2005gi}. 
They need to be
combined with the corresponding decay channels. For a light Higgs boson,
like one with $120 \GeV$ as we consider here, the main decay mode is
into a pair of bottom quarks. Also decays via off-shell $W$ and $Z$
pairs lead to observable channels.
Decays into taus can only be combined with weak-boson
fusion~\cite{wbf_tau}, as we need to reconstruct the invariant mass of
the tau pair, which is only known in the collinear limit~\cite{coll_taus}.
This channel is one of the discovery modes for a light Higgs and also
allows us to determine its mass with a precision of $\sim 5 \GeV$.
The decay into photons is loop-induced and therefore only has a small
branching fraction of $\sim 0.26 \%$ for a $120 \GeV$ Higgs.
Nevertheless the flat background, which is well subtractable by a
side-band analysis, and a good $\gamma\gamma$ mass
resolution make this mode a discovery channel and allow us to measure
the mass with a precision $\mathcal{O}(100\;\text{MeV})$.

In total we obtain the channels described in Table~\ref{tab:channels}.
The numerical values for Higgs production we get from
Ref.~\cite{spira_hqq} and
for the decay from a modified version of HDECAY~\cite{hdecay}.
We do not include any channel which will only be measured at a later
stage like the second-generation fermions. Such channels typically
determine one additional parameter and therefore do not feed back into
the analysis here. This also includes the Higgs self-couplings, which
are important to establish the nature of electro-weak symmetry breaking,
but are notoriously hard to determine
experimentally~\cite{selfcoup,quartic}.

\section{Subjet Techniques}

In this section we will discuss in more detail the recent development of
jet substructure techniques. Its presentation is based on the
original paper~\cite{subjet}. We consider associated production with
vector-bosons with Higgs decays into a bottom-quark pair. A standard
analysis would be overwhelmed by the large QCD background. Therefore we
only consider a regime where both bosons are back-to-back and have
large transverse momenta. For transverse momenta of the Higgs larger
than $200 \GeV$ only about one in twenty events survives. This is
compensated by several advantages. The boost leads to central decay
products, which can be tagged more easily. Also on-shell top quarks
cannot simulate this kind of behavior and are no longer a background.
Finally, the channel with $Z$ decaying into neutrinos becomes visible as
large missing transverse energy.

\begin{figure}
\includegraphics[width=\textwidth]{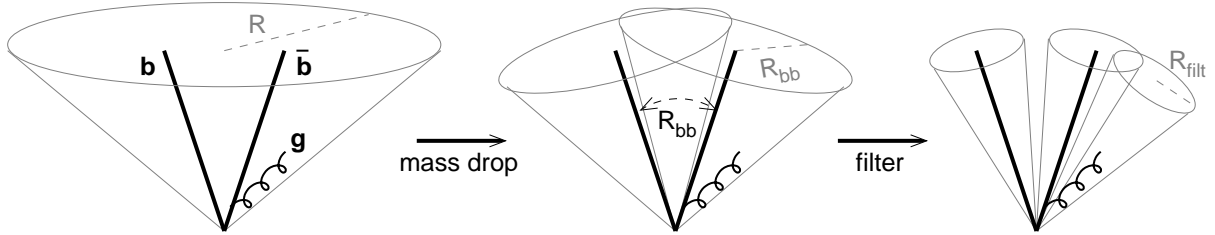}
\caption{Illustration of the subjet algorithm. Starting with a fat jet,
we first undo the last stages of the jet clustering until we have
identified the bottom jets. We then apply a filter to remove the
underlying event while retaining hard radiation from the Higgs decay
products.
Figure taken from Ref.~\protect\cite{subjet}.}
\label{fig:subjet}
\end{figure}

Such a boost also means that often the jet reconstruction algorithm will
not be able to resolve both bottom quarks. Instead it will combine them
into a single fat jet. For $R \simeq \frac{3 m_H}{p_T}$ this happens in
roughly $75 \%$ of the cases. Therefore we need to identify such a fat
jet. The Cambridge/Aachen jet algorithm~\cite{camac} 
has thereby shown to give the
best results for the following procedure. Starting with a high-$p_T$ jet
$j$ we first undo the last stage of clustering so that we obtain two
subjets $j_1, j_2$ with $m_{j_1} > m_{j_2}$. Next we check if there has
been a significant mass drop $\max(m_{j_1}, m_{j_2}) < 0.67 m$.
Additionally we test if the splitting is not too asymmetric $ y =
\frac{\min(p^2_{T,j_1}, p^2_{T,j_2})}{m_j^2} \Delta R^2_{j_1,j_2}
> 0.09$. If either of the conditions is not fulfilled, we take $j_1$ as
$j$ and repeat the steps.

Otherwise we check if both subjets have $b$ tags and if not, reject the
event. This candidate Higgs event is then filtered. We resolve the
structure with a finer $R$ separation $R_{\text{filt}} = \min(0.3,
R_{b\bar{b}}/2)$ and take the three hardest subjets. These are typically
the two $b$ jets and the leading radiation. A graphical description of
the algorithm is depicted in Fig.~\ref{fig:subjet}.

The ATLAS collaboration has performed a full study using these
techniques~\cite{subjetatlas} and they obtain a statistical significance of
$3.7\sigma$ for a $120 \GeV$ Higgs boson and a luminosity of $30 \ifb$.
Including $15\%$ systematic uncertainty a significance of $3.0 \sigma$
can be achieved.

This method has also been applied to associated Higgs production with top
quarks and bottom quark decays~\cite{subjettth}. Here the subjet technique can
help reduce the combinatorial background, which degrades the standard
analysis. For a luminosity of $100 \ifb$ the statistical significance of
this channel is $4.1$ standard deviations.
Additionally it has been shown that further combining different subjet
techniques can enhance the statistical significance even
more~\cite{Soper:2010xk}.

\section{Calculational Setup}

For our analysis we assume a generalization of the SM Higgs sector with
arbitrary couplings. Any coupling to particle $j$ present in the SM we
modify according to
\begin{equation}
g_{jjH} \rightarrow g_{jjH}^{\text{SM}} ( 1 + \Delta_{jjH} )
\end{equation}
where the $\Delta_{jjH}$ are independent of each other. As a global sign
change in the Higgs couplings is not observable, we can take $g_{WWH}$
to be always positive, or $\Delta_{WWH} > -1$. 
Additionally loop-induced couplings to the photon and gluon are
relevant. Here also the modified tree-level couplings enter. Also we
allow for further dimension-five operators from new physics in the
Lagrangian. An example for such a term are the additional loop
contributions from supersymmetric partners.
Therefore these couplings are modified to
\begin{equation} 
g_{jjH} \rightarrow g_{jjH}^{\text{SM}} ( 1 + \Delta_{jjH}^{\text{SM}} +
\Delta_{jjH}) \ ,
\end{equation} 
where $g_{jjH}^{\text{SM}}$ is the loop-induced coupling in the SM,
$\Delta_{jjH}^{\text{SM}}$ the contribution from modified tree-level
couplings and $\Delta_{jjH}$ the additional dimension-five part.
We also include the masses of the Higgs boson and the top- and bottom-quark
into our parameter set. The errors on their measurements are large
enough that their influence on the coupling determination should be
taken into account.

Furthermore we need to specify our treatment of the total width. This
could also receive further contributions 
\begin{equation}
\Gamma_{\text{tot}} = \Gamma_{\text{tot}}^\text{SM} ( 1 + \Delta_\Gamma)
\ , \quad \Delta_\Gamma \ge 0 \ .
\end{equation}
A simultaneous scaling of all couplings and the total width where
$g^4/\Gamma_{\text{tot}} = \text{const.} $ leaves all rates the same and
we therefore cannot distinguish this. We will hence fix the total width
to the sum of observed partial widths.
%%%, and discuss a scenario where this
%%% misleads us in section~\ref{sec:unobserved}.

The treatment of errors is a crucial part of the analysis. Statistical
errors on individual channels are from counting experiments and
therefore of Poisson type. The systematic errors are taken as fully
correlated between the channels. They are derived from large event
samples and can therefore safely assumed to be Gaussian. To combine
these two types of errors we have devised an approximate formula of
summing the inverse of the log-likelihoods~\cite{Lafaye:2009vr}
\begin{equation}
\frac{1}{\tilde\chi^2} \equiv \frac1{-2 \log L} 
  = \sum_i \frac1{-2 \log L_i} \ .
\end{equation}
For a sum of Gaussians this gives the correct result of adding the
errors in quadrature, while we have checked numerically that in all
other cases we have a very good agreement with the exact way of a
mathematical convolution.
For theory errors we use the RFit prescription~\cite{ckmfitter}. For a deviation
smaller than the theory error, the partial log-likelihood of the
measurement is zero, while outside of this region it falls off with the
combined experimental error. 
For the assumed numerical values of the errors we refer to
Ref.~\cite{Lafaye:2009vr}.

\section{Results}
\subsection{Likelihood map}
\begin{figure}
\psfrag{HWW}[c][][1][0]{ \raisebox{0.5em}{$\Delta_{WWH}$}}
\psfrag{HZZ}[c][][1][0]{ \raisebox{0.5em}{$\Delta_{ZZH}$}}
\psfrag{Htt}[c][][1][0]{ \raisebox{0.5em}{$\Delta_{ttH}$}}
\psfrag{Hbb}[c][][1][0]{ \raisebox{0.5em}{$\Delta_{bbH}$}}
\psfrag{Htautau}[c][][1][0]{ \raisebox{0.5em}{$\Delta_{\tau\tau H}$}}
\psfrag{Hgg}[c][][1][0]{ \raisebox{0.5em}{$\Delta_{ggH}$}}
\psfrag{Hgamgam}[c][][1][0]{ \raisebox{0.5em}{$\Delta_{\gamma\gamma H}$}}
\psfrag{Hii}[c][][1][0]{ \raisebox{0.5em}{$\Delta_\Gamma$}}
\psfrag{-9}[c][c][1][0]{-9}
\psfrag{-5}[c][c][1][0]{-5}
\psfrag{-4}[c][c][1][0]{-4}
\psfrag{-3}[c][c][1][0]{-3}
\psfrag{-2}[c][c][1][0]{-2}
\psfrag{-1}[c][c][1][0]{-1}
\psfrag{ 0}[c][c][1][0]{0}
\psfrag{ 0.0005}[c][c][1][0]{0.0005 }
\psfrag{ 0.001}[c][c][1][0]{0.001 }
\psfrag{ 0.002}[c][c][1][0]{0.002 }
\psfrag{ 0.003}[c][c][1][0]{0.003 }
\psfrag{ 0.004}[c][c][1][0]{0.004 }
\psfrag{ 0.006}[c][c][1][0]{0.006 }
\psfrag{ 0.008}[c][c][1][0]{0.008 }
\psfrag{-0.02}[c][c][1][0]{-0.02 }
\psfrag{ 0.02}[c][c][1][0]{0.02 }
\psfrag{ 0.04}[c][c][1][0]{0.04 }
\psfrag{ 0.06}[c][c][1][0]{0.06 }
\psfrag{ 0.08}[c][c][1][0]{0.08 }
\psfrag{ 0.1}[c][c][1][0]{0.1 }
\psfrag{ 0.2}[c][c][1][0]{0.2 }
\psfrag{ 0.4}[c][c][1][0]{0.4 }
\psfrag{ 0.6}[c][c][1][0]{0.6 }
\psfrag{ 0.8}[c][c][1][0]{0.8 }
\psfrag{ 1}[c][c][1][0]{1}
\psfrag{ 2}[c][c][1][0]{2}
\psfrag{ 3}[c][c][1][0]{3}
\psfrag{ 4}[c][c][1][0]{4}
\psfrag{ 5}[c][c][1][0]{5}
\psfrag{ 6}[c][c][1][0]{6}
\psfrag{ 7}[c][c][1][0]{7}
\psfrag{ 8}[c][c][1][0]{8}
\psfrag{ 9}[c][c][1][0]{9}
\psfrag{ 10}[c][c][1][0]{10}
\psfrag{ 15}[c][c][1][0]{15}
\psfrag{ 20}[c][c][1][0]{20}
\psfrag{ 160}[c][c][1][0]{160 }
\psfrag{ 180}[c][c][1][0]{180 }
\psfrag{ 200}[c][c][1][0]{200 }
\psfrag{TOPm}[c][c][1][0]{$m_t$}
\psfrag{TOPt}[c][c][1][0]{}
\psfrag{m}[c][c][1][0]{$m$}
{\raggedright $1/\Delta \chi^2$\\[2ex]}
\includegraphics[width=0.15\textwidth]{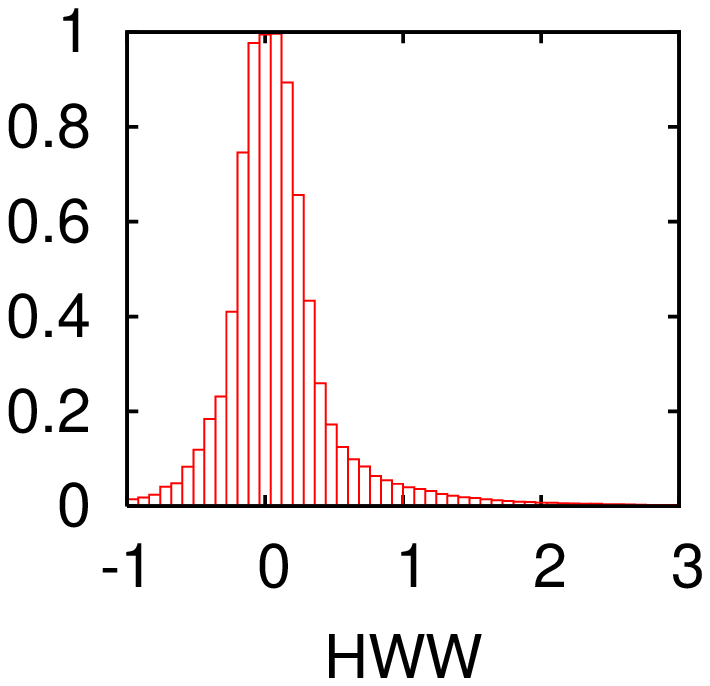} 
\includegraphics[width=0.15\textwidth]{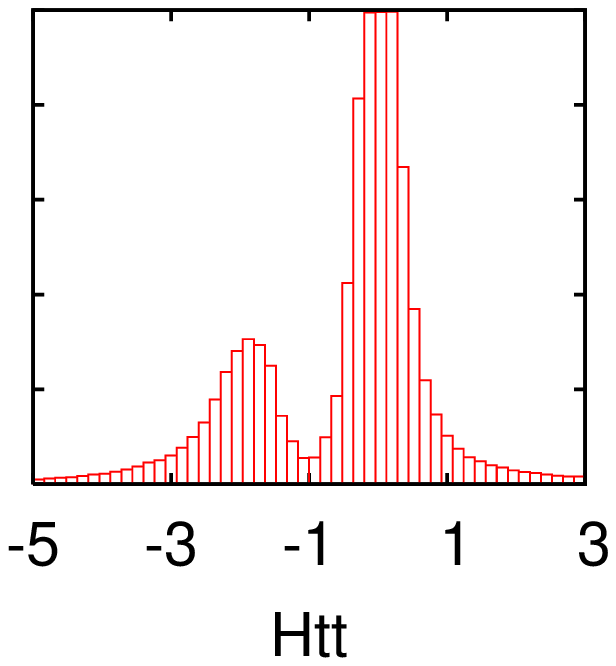} 
\includegraphics[width=0.15\textwidth]{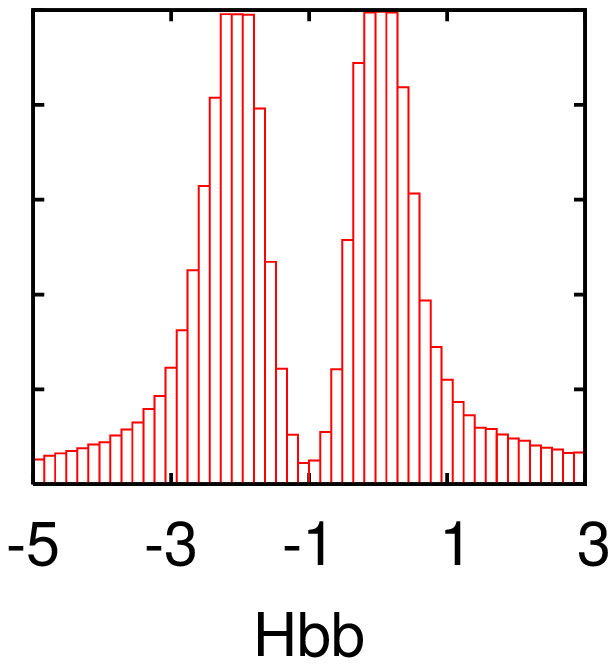}
\hspace*{2em}
\includegraphics[width=0.15\textwidth]{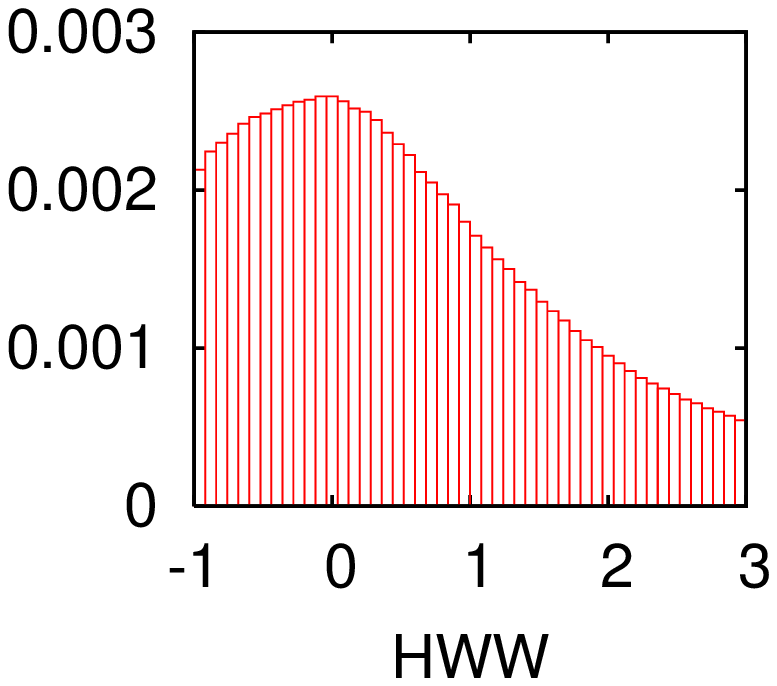} 
\includegraphics[width=0.15\textwidth]{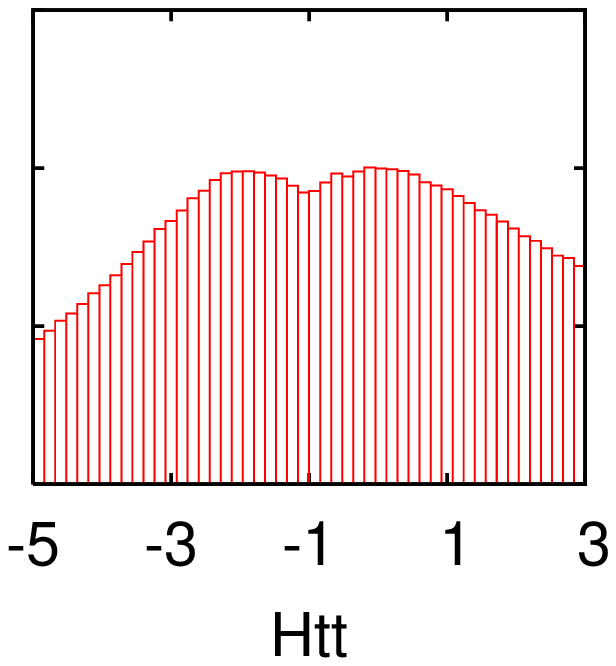} 
\includegraphics[width=0.15\textwidth]{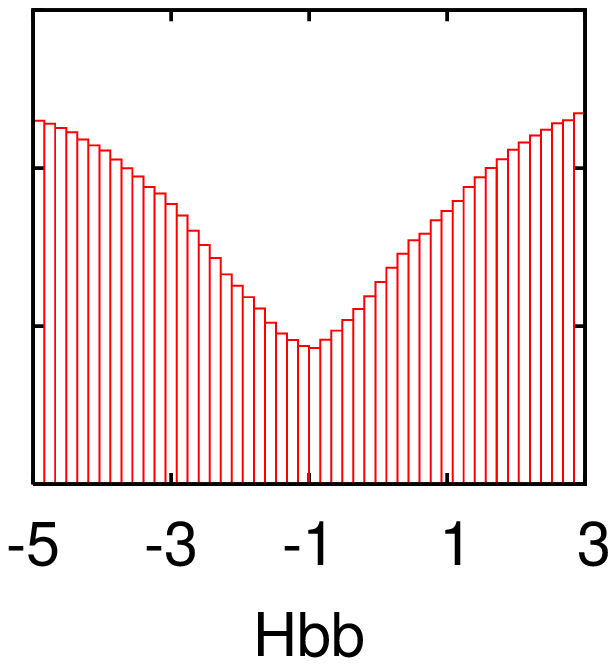} \\
\includegraphics[width=0.15\textwidth]{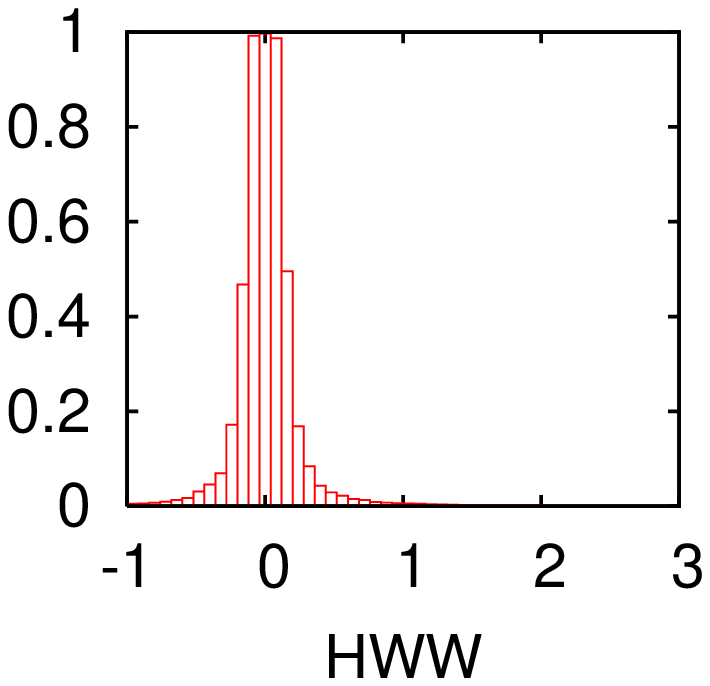} 
\includegraphics[width=0.15\textwidth]{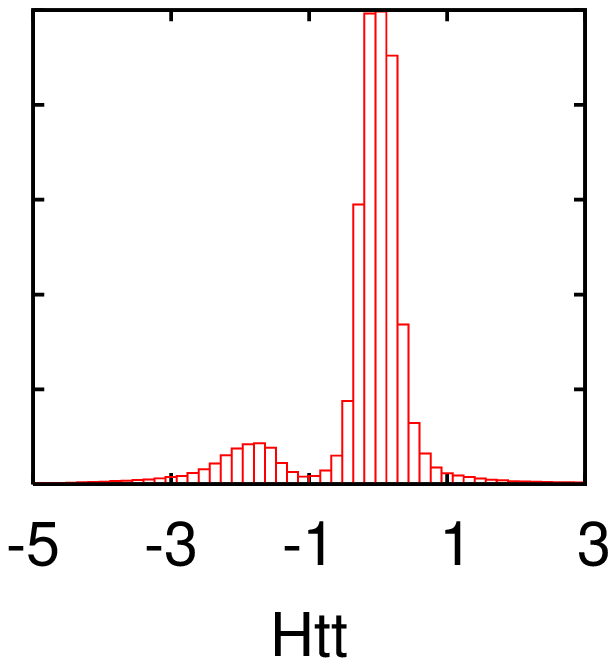} 
\includegraphics[width=0.15\textwidth]{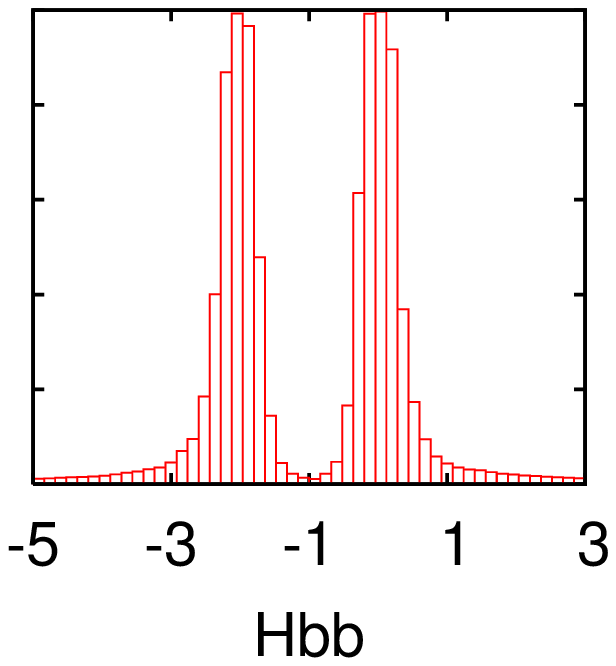} 
\hspace*{2em}
\includegraphics[width=0.15\textwidth]{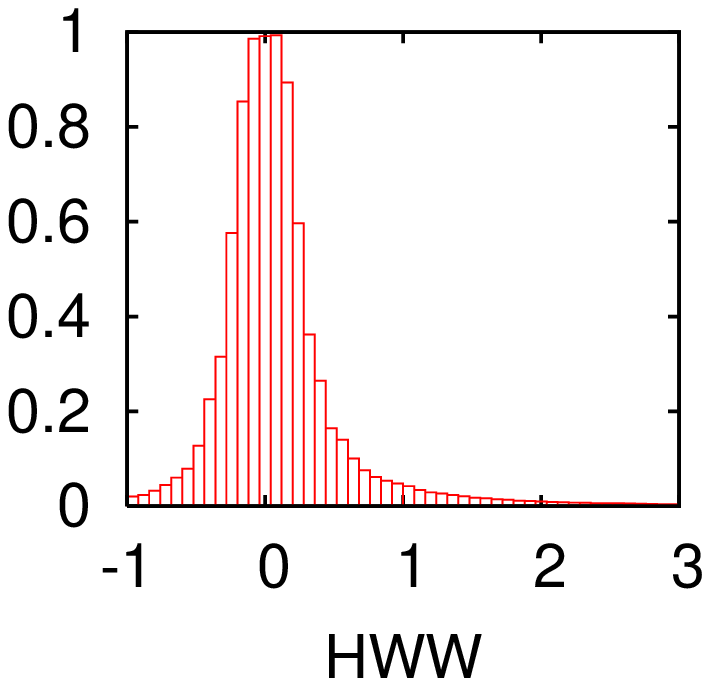} 
\includegraphics[width=0.15\textwidth]{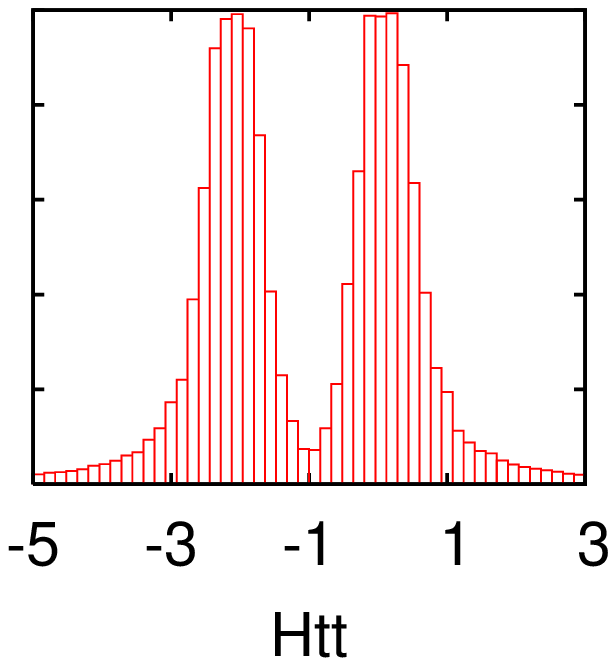} 
\includegraphics[width=0.15\textwidth]{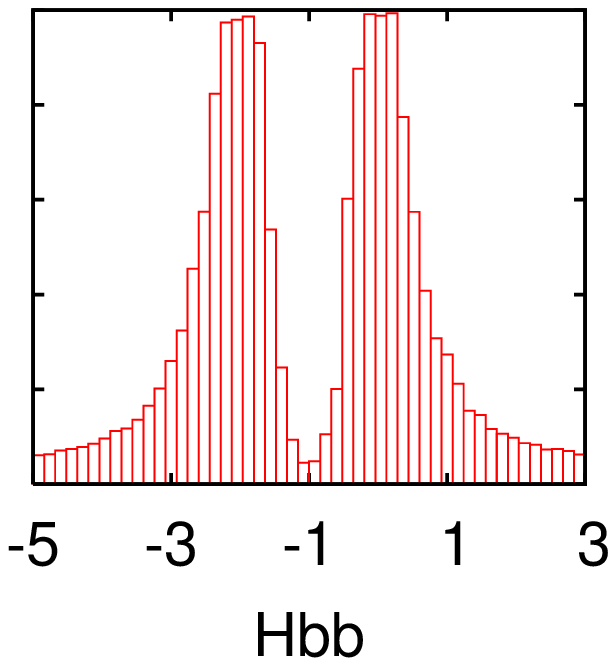} \\
\includegraphics[width=0.15\textwidth]{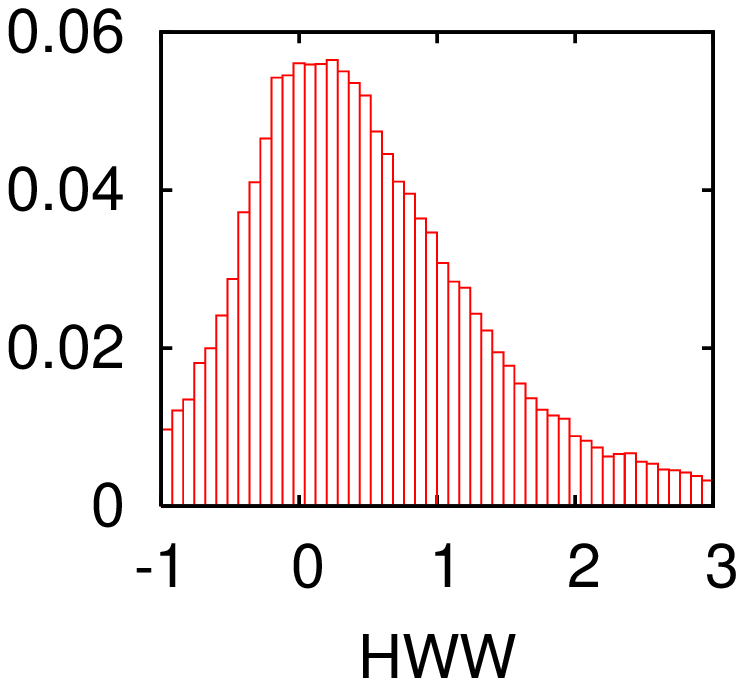} 
\includegraphics[width=0.15\textwidth]{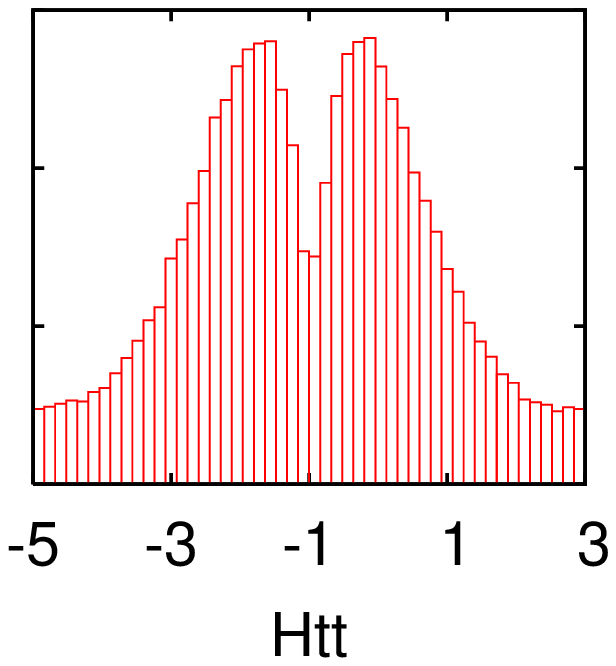} 
\includegraphics[width=0.15\textwidth]{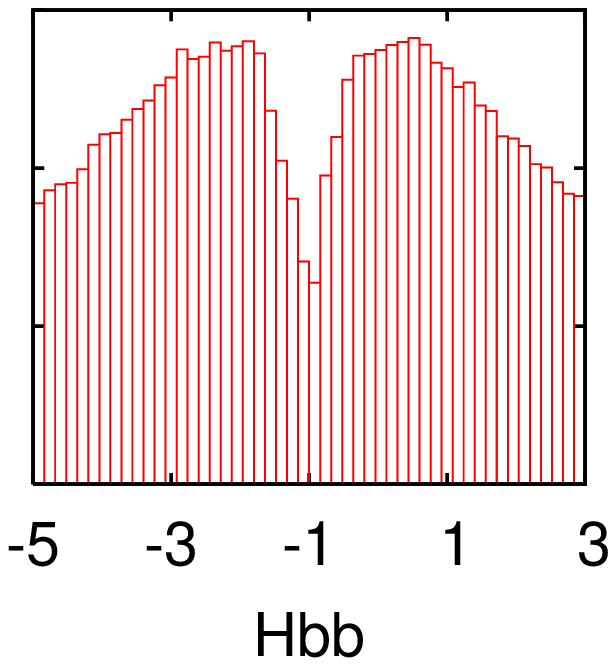} 
\caption{Distributions for different couplings. Starting from the top
left set with profile likelihoods for an unsmeared data set with an
integrated luminosity of $30 \ifb$ and no effective couplings, we modify
these in the other sets: Bayesian marginalization (top right), $300
\ifb$ (middle left), including effective coupling (middle right), and
smeared data set with effective couplings (bottom).
Plots taken from Ref.~\protect\cite{Lafaye:2009vr}.}
\label{fig:smplots}
\end{figure}

In Fig.~\ref{fig:smplots} we show distributions for the three couplings
$\Delta_{WWH}$, $\Delta_{ttH}$ and $\Delta_{bbH}$. In the left top row
we show profile likelihoods for an unsmeared data set with an integrated
luminosity of $30 \ifb$. Additional contributions to the loop-induced
couplings we set to zero here. For all couplings we see a peak at the
correct solution of $\Delta = 0$. For both $ttH$ and $bbH$ a second peak
at $\Delta=-2$ appears, which corresponds to a flipped sign of the
coupling. For the top-quark the likelihood of this solution is strongly
reduced. The interference with a $W$-loop in the effective Higgs-photon
coupling allows us to determine the correct sign. In principle the same
effect would also be true for the bottom quark. Its contribution to the
effective photon or gluon coupling, which would fix the sign with
respect to $ttH$, is too small to see an effect.

On the right-hand side we take the same input in the top row, but now
use Bayesian marginalization. The interference effect, which breaks the
sign degeneracy in $ttH$, is completely washed out by volume effects. 
The $bbH$ coupling shows a peculiar effect. Large values of the coupling
are more likely than the correct one. The branching ratio into bottom
quarks is much less affected by changing the coupling than other ones,
because this also increases the total width significantly and the effect
partly cancels. On the other hand large values allow larger changes in
the other couplings without relevant changes of the overall rates.
Therefore we have more parameter space available, so this is a pure
volume effect.
To understand the correlations we see Bayesian probabilities are less
useful here and therefore we will show only profile likelihoods from now
on. 

On the left-hand side of the middle row of Figure~\ref{fig:smplots} we
show a ten-fold increase of luminosity to $300 \ifb$ compared to the top
row. The general features stay unchanged, while the errors go down
significantly, because the precision is primarily statistically limited.
On the right-hand side we allow for additional contributions to the
$ggH$ and $\gamma\gamma H$ couplings at $30 \ifb$. We see that both
peaks for the top-quark coupling now have the same height.  The
effective photon coupling can no longer break the sign degeneracy of
$ttH$. Any mismatch is compensated by appropriately dialing the new
term.

In the bottom row we move from the true data set to a smeared one. This
we have obtained by randomly smearing each measurement according to its
errors. We see that the overall behavior does not change significantly.
The best-fitting points move slightly away from their true values. Also
the peak structure gets broader, as different measurements try to pull
the parameters into different directions.

\subsection{Error determination}
\begin{table}
\caption{Errors on the measurements form 10000 toy experiments. We quote
absolute errors on the couplings for $30 \ifb$, where additional
contributions to the $ggH$ and $\gamma\gamma H$ couplings are either
forbidden or allowed. For the latter we also show errors on the 
ratio of the coupling to $WWH$. Table taken from
Ref.~\protect\cite{Lafaye:2009vr}.}
~\\
\begin{tabular}{|@{}l@{}|@{\ }l@{\ }|@{\ }l@{\ }l@{\ }|@{\ }l@{\ }|@{\ }l@{\ }l@{\ }|@{\ }l@{\ }|@{\ }l@{\ }l@{\ }|}
\hline
 &         \multicolumn{3}{c|@{\ }}{no eff. couplings} &
           \multicolumn{3}{c|@{\ }}{with eff. couplings} &
           \multicolumn{3}{c|}{ratio $\Delta_{jjH/WWH}$} \\
 & $\sigma_\mathrm{symm}$ & $\sigma_\mathrm{neg}$ & $\sigma_\mathrm{pos}$
 & $\sigma_\mathrm{symm}$ & $\sigma_\mathrm{neg}$ & $\sigma_\mathrm{pos}$
 & $\sigma_\mathrm{symm}$ & $\sigma_\mathrm{neg}$ & $\sigma_\mathrm{pos}$
\\\hline 
$\Delta_{WWH}$             
 & $\pm\,0.23$ & $-\,0.21$ & $+\,0.26$ 
 & $\pm\,0.24$ & $-\,0.21$ & $+\,0.27$ 
 &\phantom{$\pm$} --- &\phantom{$-$} ---
&\phantom{$+$} --- \\
$\Delta_{ZZH}$             
 & $\pm\,0.36$ & $-\,0.40$ & $+\,0.35$ 
 & $\pm\,0.31$ & $-\,0.35$ & $+\,0.29$ 
 & $\pm\,0.41$ & $-\,0.40$ & $+\,0.41$ \\
$\Delta_{ttH}$       
 & $\pm\,0.41$ & $-\,0.37$ & $+\,0.45$ 
 & $\pm\,0.53$ & $-\,0.65$ & $+\,0.43$ 
 & $\pm\,0.51$ & $-\,0.54$ & $+\,0.48$ \\
$\Delta_{bbH}$       
 & $\pm\,0.45$ & $-\,0.33$ & $+\,0.56$ 
 & $\pm\,0.44$ & $-\,0.30$ & $+\,0.59$ 
 & $\pm\,0.31$ & $-\,0.24$ & $+\,0.38$ \\
$\Delta_{\tau\tau{}H}$ 
 & $\pm\,0.33$ & $-\,0.21$ & $+\,0.46$ 
 & $\pm\,0.31$ & $-\,0.19$ & $+\,0.46$ 
 & $\pm\,0.28$ & $-\,0.16$ & $+\,0.40$ \\
$\Delta_{\gamma\gamma{}H}$ & 
 \phantom{$\pm$} --- &\phantom{$\pm$} --- &\phantom{$-$} ---
 & $\pm\,0.31$ & $-\,0.30$ & $+\,0.33$
 & $\pm\,0.30$ & $-\,0.27$ & $+\,0.33$  \\
$\Delta_{ggH}$             & 
 \phantom{$\pm$} --- &\phantom{$\pm$} --- &\phantom{$-$} ---
 & $\pm\,0.61$ & $-\,0.59$ & $+\,0.62$ 
 & $\pm\,0.61$ & $-\,0.71$ & $+\,0.46$  \\
$m_H$                      
 & $\pm\,0.26$ & $-\,0.26$ & $+\,0.26$ 
 & $\pm\,0.25$ & $-\,0.26$ & $+\,0.25$  
 & \phantom{$\pm$} --- &\phantom{$\pm$} --- &\phantom{$-$} ---
 \\
$m_b$                      
 & $\pm\,0.071$& $-\,0.071$& $+\,0.071$
 & $\pm\,0.071$& $-\,0.071$& $+\,0.072$  
 & \phantom{$\pm$} --- &\phantom{$\pm$} --- &\phantom{$-$} ---
 \\
$m_t$                      
 & $\pm\,1.00$ & $-\,1.03$ & $+\,0.98$ 
 & $\pm\,0.99$ & $-\,1.00$ & $+\,0.98$
 & \phantom{$\pm$} --- &\phantom{$\pm$} --- &\phantom{$-$} ---
 \\ \hline
\end{tabular}
\label{tab:errors}
\end{table}

To determine the errors we perform 10000 toy experiments, where we have
smeared each measurement around the true data point including all
experimental and theory errors. The resulting distribution we then fit
with a Gaussian and extract $\sigma_\mathrm{symm}$. As the errors are
not necessarily symmetric, we also fit two-half Gaussians with the same
maximum and same height at the maximum, but different errors below
($\sigma_\mathrm{neg}$) and above ($\sigma_\mathrm{pos}$) the maximum.
The fit-region is in all cases the central part within one standard
deviation. 

From the results in Table~\ref{tab:errors} we see that the mass
measurements hardly play any role. Their distributions are completely
symmetric Gaussians and the errors correspond to the input values.
The $WWH$ coupling is the most precisely determined one. Its value does
not change once we allow effective couplings, which means that the
indirect determination via the loop-induced $g_{\gamma\gamma H}$ is not
important for a precise determination. The situation is completely
different for $ttH$. As main contribution to $g_{ggH}$ and sub-leading
to $g_{\gamma\gamma H}$ the effective couplings significantly increase
the error. Both main measurements for $\tau\tau H$ and $bbH$ are linked
with a production mode where the $WWH$ coupling enters, namely
vector-boson fusion and associated production with subjet techniques,
respectively. Therefore effective couplings do not change the errors
here. 
The error on $ZZH$ shows a particular effect. It decreases once we
include effective couplings. This is an effect of the correlations
between measurements. The main determination channel is gluon-fusion
production with Higgs decay into a pair of $Z$ bosons. Without effective
couplings the production side is linked to the top-quark coupling, which
is constrained by many other channels. Including them, this connection
is removed and the error on the measurement need not be compensated by
changing $ZZH$ alone. Instead both can be changed, and due to the
positive correlation between the two couplings this reduces the error.

One might expect that forming ratios of coupling constants could improve
the precision, as always certain combinations appear in our rate
measurements. Due to its relatively small error the $WWH$ coupling
serves well as base value.
Hence we also define a deviation on the coupling ratio
to the $WWH$ coupling as
\begin{equation}
\frac{g_{jjH}}{g_{WWH}} \rightarrow \left( \frac{g_{jjH}}{g_{WWH}}
\right)^{\text{SM}} \left( 1 + \Delta_{jjH/WWH} \right) \ .
\end{equation}
The results we show in the right column of Table~\ref{tab:errors}. In
particular $bbH$ benefits from forming the ratio. This is due to the
total width, which appears in all measurements. The $bbH$ coupling
yields the largest contribution for a light Higgs boson and this leads
to strong correlations to all other couplings.

\section{Conclusions}
In this article we have studied the determination of Higgs couplings at
the LHC. As model we have assumed the Standard Model with free Higgs
couplings, so that we are independent from the exact realization of new
physics, if any. Using SFitter as a tool we map a set of LHC
measurements onto the multi-dimensional parameter space. We have taken
effects from statistical, correlated systematic and box-shaped theory
errors into account. 

We find that we can determine the couplings with a precision of $20 - 40
\%$. 
The improved accuracy from the newly developed subjet techniques is
thereby an important ingredient. It helps to determine the bottom-quark
Higgs coupling, which influences all others via its contribution to the
total width.

\section*{Acknowledgments}
We would like to thank the organizers of ``Rencontres de Moriond EW
2010'' for the inspiring atmosphere during the workshop and for
financial support. We acknowledge support by the Deutsche
Forschungsgemeinschaft via the Sonderforschungsbereich/Transregio
SFB/TR-9 “Computational Particle Physics” and the Initiative and
Networking Fund of the Helmholtz Association, contract HA-101 (“Physics
at the Terascale”).


\section*{References}
\begin{thebibliography}{99}

\bibitem{higgs}
 P.~W.~Higgs,
  %``Broken symmetries, massless particles and gauge fields,''
  Phys.\ Lett.\  {\bf 12}, 132 (1964);
  %%CITATION = PHLTA,12,132;%%
 P.~W.~Higgs,
  %``BROKEN SYMMETRIES AND THE MASSES OF GAUGE BOSONS,''
  Phys.\ Rev.\ Lett.\  {\bf 13}, 508 (1964);
  %%CITATION = PRLTA,13,508;%%
 F.~Englert and R.~Brout,
  %``BROKEN SYMMETRY AND THE MASS OF GAUGE VECTOR MESONS,''
  Phys.\ Rev.\ Lett.\  {\bf 13}, 321 (1964).
  %%CITATION = PRLTA,13,321;%%

\bibitem{reviews}
 A.~Djouadi,
  %``The anatomy of electro-weak symmetry breaking. I: The Higgs boson
  %in  the
  %standard model,''
  Phys.\ Rept.\  {\bf 457}, 1 (2008)
  %[arXiv:hep-ph/0503172].
  %%CITATION = PRPLC,457,1;%%
 V.~B\"uscher and K.~Jakobs,
  %``Higgs boson searches at hadron colliders,''
  Int.\ J.\ Mod.\ Phys.\  A {\bf 20}, 2523 (2005);
  %[arXiv:hep-ph/0504099].
  %%CITATION = IMPAE,A20,2523;%%
 D.~Rainwater,
  %``Searching for the Higgs boson,''
  arXiv:hep-ph/0702124.
  %%CITATION = HEP-PH/0702124;%%

\bibitem{Alcaraz:2007ri}
  J.~Alcaraz {\it et al.}  [LEP Collaborations and ALEPH Collaboration
and
                  DELPHI Collaboration an],
  %``Precision Electroweak Measurements and Constraints on the Standard
  %Model,''
  arXiv:0712.0929 [hep-ex].
  %%CITATION = ARXIV:0712.0929;%%

\bibitem{Collaboration:2008ub}
    [ALEPH Collaboration and CDF Collaboration and D0 Collaboration and
an],
  %``Precision Electroweak Measurements and Constraints on the Standard
  %Model,''
  arXiv:0811.4682 [hep-ex].
  %%CITATION = ARXIV:0811.4682;%%

\bibitem{duehrssen}
 M.~D\"uhrssen, S.~Heinemeyer, H.~Logan, D.~Rainwater, G.~Weiglein and
D.~Zeppenfeld,
  %``Extracting Higgs boson couplings from LHC data,''
  Phys.\ Rev.\ D {\bf 70}, 113009 (2004).
  %[arXiv:hep-ph/0406323].
  %%CITATION = HEP-PH 0406323;%%

%\cite{Lafaye:2009vr}
\bibitem{Lafaye:2009vr}
  R.~Lafaye, T.~Plehn, M.~Rauch, D.~Zerwas and M.~Duhrssen,
  %``Measuring the Higgs Sector,''
  JHEP {\bf 0908}, 009 (2009)
  [arXiv:0904.3866 [hep-ph]].
  %%CITATION = JHEPA,0908,009;%%

\bibitem{atlas_tdr}
  G.~Aad {\it et al.}  [The ATLAS Collaboration],
  %``Expected Performance of the ATLAS Experiment - Detector, Trigger
  %and Physics,''
  arXiv:0901.0512 [hep-ex].
  %%CITATION = ARXIV:0901.0512;%%

\bibitem{cms_tdr}
  G.~L.~Bayatian {\it et al.}  [CMS Collaboration],
  %``CMS technical design report, volume II: Physics performance,''
  J.\ Phys.\ G {\bf 34}, 995 (2007).
  %%CITATION = JPHGB,G34,995;%%

\bibitem{wbf_vertex}
 J.~R.~Dell'Aquila and C.~A.~Nelson,
  %``Distinguishing A Spin 0 Technipion And An Elementary Higgs Boson:
  %V1 V2
  %Modes With Decays Into Anti-Lepton (A) Lepton (B) And/Or Anti-Q (A) Q
  %(B),''
  Phys.\ Rev.\  D {\bf 33}, 93 (1986).;
  %%CITATION = PHRVA,D33,93;%%
 T.~Plehn, D.~Rainwater and D.~Zeppenfeld,
  %``Determining the structure of Higgs couplings at the LHC,''
  Phys.\ Rev.\ Lett.\  {\bf 88}, 051801 (2002);
  %%CITATION = HEP-PH 0105325;%%
 C.~P.~Buszello, I.~Fleck, P.~Marquard and J.~J.~van der Bij,
  %``Prospective Analysis Of Spin- And Cp-Sensitive Variables In H $\to$
  %Z Z
  %$\to$ L(1)+ L(1)- L(2)+ L(2)- At The Lhc,''
  Eur.\ Phys.\ J.\  C {\bf 32}, 209 (2004);
  %[arXiv:hep-ph/0212396].
  %%CITATION = EPHJA,C32,209;%%
 V.~Hankele, G.~Klamke, D.~Zeppenfeld and T.~Figy,
  %``Anomalous Higgs boson couplings in vector boson fusion at the CERN
  %LHC,''
  Phys.\ Rev.\  D {\bf 74}, 095001 (2006);
  %[arXiv:hep-ph/0609075].
  %%CITATION = PHRVA,D74,095001;%%
 C.~Ruwiedel, N.~Wermes and M.~Schumacher,
  %``Prospects for the measurement of the structure of the coupling of a
  %Higgs
  %boson to weak gauge bosons in weak boson fusion with the ATLAS
  %detector,''
  Eur.\ Phys.\ J.\  C {\bf 51}, 385 (2007).
  %%CITATION = EPHJA,C51,385;%%

\bibitem{susy}
 for a pedagogical introduction see e.g.$\;$
 S.~P.~Martin,
  %``A supersymmetry primer,''
  arXiv:hep-ph/9709356;
  %%CITATION = HEP-PH/9709356;%%
 I.~J.~R.~Aitchison,
  %``Supersymmetry and the MSSM: An elementary introduction,''
  arXiv:hep-ph/0505105;
  %%CITATION = HEP-PH/0505105;%%
 J.~F.~Gunion and H.~E.~Haber,
  %``The CP-conserving two-Higgs-doublet model: The approach to the
  %decoupling
  %limit,''
  Phys.\ Rev.\  D {\bf 67}, 075019 (2003).
  %[arXiv:hep-ph/0207010].
  %%CITATION = PHRVA,D67,075019;%%
\bibitem{m_h}
 H.~E.~Haber, R.~Hempfling and A.~H.~Hoang,
  %``Approximating the radiatively corrected Higgs mass in the minimal
  %supersymmetric model,'' 
  Z.\ Phys.\ C {\bf 75}, 539 (1997);
  %[arXiv:hep-ph/9609331]. 
  %%CITATION = HEP-PH 9609331;%%
 G.~Degrassi, S.~Heinemeyer, W.~Hollik, P.~Slavich and G.~Weiglein,
  %``Towards high-precision predictions for the MSSM Higgs sector,''
  Eur.\ Phys.\ J.\  C {\bf 28}, 133 (2003);
  %[arXiv:hep-ph/0212020].
  %%CITATION = EPHJA,C28,133;%%
 T.~Hahn, S.~Heinemeyer, W.~Hollik, H.~Rzehak, G.~Weiglein and
K.~Williams,
  %``Higher-order corrected Higgs bosons in FeynHiggs 2.5,''
  Pramana {\bf 69}, 861 (2007).
  %[arXiv:hep-ph/0611373].
  %%CITATION = PRAMC,69,861;%%

%\cite{Espinosa:2010vn}
\bibitem{Espinosa:2010vn}
  J.~R.~Espinosa, C.~Grojean and M.~Muhlleitner,
  %``Composite Higgs Search at the LHC,''
  arXiv:1003.3251 [hep-ph].
  %%CITATION = ARXIV:1003.3251;%%

\bibitem{ckmfitter}
 A.~H\"ocker, H.~Lacker, S.~Laplace and F.~Le Diberder,
  %``A new approach to a global fit of the CKM matrix,''
  Eur.\ Phys.\ J.\  C {\bf 21}, 225 (2001);
  %[arXiv:hep-ph/0104062].
  %%CITATION = EPHJA,C21,225;%%
 J.~Charles, A.~H\"ocker, H.~Lacker, F.~R.~Le Diberder and S.~T'Jampens,
  %``Bayesian statistics at work: The troublesome extraction of the CKM
  %phase
  %alpha,''
  arXiv:hep-ph/0607246.
  %%CITATION = HEP-PH/0607246;%%

\bibitem{sfitter}
 R.~Lafaye, T.~Plehn, M.~Rauch and D.~Zerwas,
  %``Measuring Supersymmetry,''
  Eur.\ Phys.\ J.\  C {\bf 54}, 617 (2008);
  %[arXiv:0709.3985 [hep-ph]].
  %%CITATION = EPHJA,C54,617;%% \\
 for earlier versions of SFitter see:
 R.~Lafaye, T.~Plehn and D.~Zerwas,
  %``SFITTER: SUSY parameter analysis at LHC and LC,''
  arXiv:hep-ph/0404282
  %%CITATION = HEP-PH 0404282;%%
 and
 %R.~Lafaye, T.~Plehn and D.~Zerwas,
  %``SFITTER: SUSY parameter determination,''
  arXiv:hep-ph/0512028.
  %%CITATION = ECONF, C0508141,ALCPG0607;%%

\bibitem{subjet}
 J.~M.~Butterworth, A.~R.~Davison, M.~Rubin and G.~P.~Salam,
  %``Jet substructure as a new Higgs search channel at the LHC,''
  Phys.\ Rev.\ Lett.\  {\bf 100}, 242001 (2008).
  %[arXiv:0802.2470 [hep-ph]].
  %%CITATION = PRLTA,100,242001;%%

\bibitem{duehrssennote}
 M.~D\"uhrssen,
 %``Prospects for the measurement of Higgs boson
 %coupling parameters in the mass range
 %from 110 - 190 GeV/c2,''
 ATL-PHYS-2003-030.

%\cite{Djouadi:2005gi}
\bibitem{Djouadi:2005gi}
  for an overview see e.g.\ 
  A.~Djouadi,
  %``The Anatomy of electro-weak symmetry breaking. I: The Higgs boson in the
  %standard model,''
  Phys.\ Rept.\  {\bf 457}, 1 (2008)
  [arXiv:hep-ph/0503172].
  %%CITATION = PRPLC,457,1;%%

\bibitem{wbf_tau}
 D.~Rainwater, D.~Zeppenfeld and K.~Hagiwara,
  %``Searching for H $\to$ tau tau in weak boson fusion at the LHC,''
  Phys.\ Rev.\ D {\bf 59}, 014037 (1999);
  %%CITATION = HEP-PH 9808468;%%
 T.~Plehn, D.~L.~Rainwater and D.~Zeppenfeld,
  %``A method for identifying H $\to$ tau tau $\to$ e+- mu-+ missing p(T)  at the
  %CERN LHC,''
  Phys.\ Rev.\ D {\bf 61}, 093005 (2000).
  %%[arXiv:hep-ph/9911385].
  %%CITATION = HEP-PH 9911385;%%

\bibitem{coll_taus}
 R.~K.~Ellis, I.~Hinchliffe, M.~Soldate and J.~J.~van der Bij,
  % ``Higgs Decay to tau+ tau-: A Possible Signature of Intermediate Mass Higgs
  %Bosons at the SSC,''
  Nucl.\ Phys.\  B {\bf 297}, 221 (1988).
  %%CITATION = NUPHA,B297,221;%%

\bibitem{spira_hqq}
  M.~Spira,
  %``HIGLU: A Program for the Calculation of the Total Higgs Production
  %Cross
  %Section at Hadron Colliders via Gluon Fusion including QCD
  %Corrections,''
  arXiv:hep-ph/9510347.
  %%CITATION = HEP-PH/9510347;%%

\bibitem{hdecay}
  A.~Djouadi, J.~Kalinowski and M.~Spira,
  % ``HDECAY: A program for Higgs boson decays in the standard model and
  % its
  %supersymmetric extension,''
  Comput.\ Phys.\ Commun.\  {\bf 108}, 56 (1998).
  %[arXiv:hep-ph/9704448].
  %%CITATION = CPHCB,108,56;%%
 
\bibitem{selfcoup}
 U.~Baur, T.~Plehn and D.~L.~Rainwater,
  %``Measuring the Higgs boson self coupling at the LHC and finite top
  %mass
  %matrix elements,''
  Phys.\ Rev.\ Lett.\  {\bf 89}, 151801 (2002)
  %[arXiv:hep-ph/0206024].
  %%CITATION = HEP-PH 0206024;%%
 and
 %U.~Baur, T.~Plehn and D.~L.~Rainwater,
  %``Determining the Higgs boson self coupling at hadron colliders,''
  Phys.\ Rev.\ D {\bf 67}, 033003 (2003);
  %[arXiv:hep-ph/0211224].
  %%CITATION = HEP-PH 0211224;%%
 A.~Dahlhoff,
  %``Prospects to measure the Higgs boson properties in ATLAS,''
  arXiv:hep-ex/0505022;
  %%CITATION = HEP-EX 0505022;%%
 for a different point of view see also
 F.~Gianotti {\it et al.},
  %``Physics potential and experimental challenges of the LHC luminosity
  %upgrade,''
  arXiv:hep-ph/0204087.
  %%CITATION = HEP-PH 0204087;%%

\bibitem{quartic}
 T.~Plehn and M.~Rauch,
   %``The quartic Higgs coupling at hadron colliders,''
   Phys.\ Rev.\ D {\bf 72}, 053008 (2005);
   %%[arXiv:hep-ph/0507321].
   %%CITATION = HEP-PH 0507321;%%
 T.~Binoth, S.~Karg, N.~Kauer and R.~R\"uckl,
  %``Multi-Higgs boson production in the standard model and beyond,''
  Phys.\ Rev.\  D {\bf 74}, 113008 (2006).
  %[arXiv:hep-ph/0608057].
  %%CITATION = PHRVA,D74,113008;%%

\bibitem{camac}
%\cite{Dokshitzer:1997in}
%\bibitem{Dokshitzer:1997in}
  Y.~L.~Dokshitzer, G.~D.~Leder, S.~Moretti and B.~R.~Webber,
  %``Better Jet Clustering Algorithms,''
  JHEP {\bf 9708}, 001 (1997)
  [arXiv:hep-ph/9707323];
  %%CITATION = JHEPA,9708,001;%%
%\cite{Wobisch:1998wt}
%\bibitem{Wobisch:1998wt}
  M.~Wobisch and T.~Wengler,
  %``Hadronization corrections to jet cross sections in deep-inelastic
  %scattering,''
  arXiv:hep-ph/9907280.
  %%CITATION = HEP-PH/9907280;%%

\bibitem{subjetatlas}
  ATLAS~Collaboration,
  ATL-PHYS-PUB-2009-088

\bibitem{subjettth}
%\cite{Plehn:2009rk}
%\bibitem{Plehn:2009rk}
  T.~Plehn, G.~P.~Salam and M.~Spannowsky,
  %``Fat Jets for a Light Higgs,''
  Phys.\ Rev.\ Lett.\  {\bf 104}, 111801 (2010)
  [arXiv:0910.5472 [hep-ph]].
  %%CITATION = PRLTA,104,111801;%%

%\cite{Soper:2010xk}
\bibitem{Soper:2010xk}
  D.~E.~Soper and M.~Spannowsky,
  %``Combining subjet algorithms to enhance ZH detection at the LHC,''
  arXiv:1005.0417 [hep-ph].
  %%CITATION = ARXIV:1005.0417;%%

\end{thebibliography}
\end{document}

